\documentclass[pra,prl,twocolumn,superscriptaddress,floatfix,showpacs,10pt]{revtex4}

\usepackage{amsfonts}
\usepackage{amsmath}
\usepackage{amssymb}
\usepackage{graphicx}

\begin{document}

\title{Hidden Messenger Revealed in Hawking Radiation:\\
a Resolution to the Paradox of Black Hole Information Loss}
\author{Baocheng Zhang}
\affiliation{State Key Laboratory of Magnetic Resonances and Atomic
and Molecular Physics, Wuhan Institute of Physics and Mathematics,
Chinese Academy of Sciences, Wuhan 430071, China}
\affiliation{Graduate University of Chinese Academy of Sciences,
Beijing 100049, China}
\author{Qing-yu Cai}
\email{qycai@wipm.ac.cn}
\affiliation{State Key Laboratory of
Magnetic Resonances and Atomic and Molecular Physics, Wuhan
Institute of Physics and Mathematics, Chinese Academy of Sciences,
Wuhan 430071, China}
\author{Li You}
\affiliation{Center for Cold Atom Physics, Chinese Academy of
Sciences, Wuhan 430071, China }
\author{Ming-sheng Zhan}
\affiliation{State Key Laboratory of Magnetic Resonances and Atomic
and Molecular Physics, Wuhan Institute of Physics and Mathematics,
Chinese Academy of Sciences, Wuhan 430071, China}
\affiliation{Center for Cold Atom Physics,
Chinese Academy of Sciences, Wuhan 430071, China }

\begin{abstract}
Using standard statistical method, we discover the existence of
correlations among Hawking radiations (of tunneled particles) from a black hole.
The information carried by such correlations is
quantified by mutual information between sequential emissions.
Through a careful counting of the entropy taken out by
the emitted particles, we show that
the black hole radiation as tunneling is
an entropy conservation process.
While information is leaked out through the radiation,
the total entropy is conserved. Thus, we conclude the
black hole evaporation process is unitary.
\end{abstract}


\maketitle


Since Hawking radiation was first discovered \cite{swh74,swh75}, its
inconsistency with quantum theory has been widely noticed.
Irrespective of what initial state a black hole starts with before
collapsing, it evolves eventually into a thermal state after being
completely exhausted into emitted radiations. Such a scenario
violates the principle of unitarity as required for quantum
mechanics and brings a serious challenge to the foundation of modern
physics. Many groups
\cite{swh76,swh05,acn87,jp92,kw89,jdb93,hm04,sl06,bp07} have
attempted addressing this puzzle of the so-called paradox of black
hole information loss. None has been successful. Most discussions
treat Hawking radiation as thermally distributed without considering
energy conservation or self-gravitation effect. Recently, Parikh and
Wilczek point out that Hawking radiation is completely non-thermal
when energy conservation is enforced \cite{pw00}. Making use of
their result, we discover the existence of non-trivial correlations
amongst Hawking radiations. A queue of correlated radiation can
transmit encoded information. By carefully counting the entropy
embedded in the sequentially emitted (tunneled out)
radiations/particles, we show that the process of Hawking radiation
is entropy conserved, contrary to entropy growth by the thermal
spectrum \cite{ZW82}. While information is carried away by Hawking
radiation, the total entropy of the black hole and the radiation is
conserved. Our work thus implies that the black hole evaporation
process, whereby Hawking radiation is emitted, is unitary.

In the past few decades, several approaches have been suggested for
resolving the paradox of black hole information loss. Hawking
initially proposed to accept information loss when quantum theory is
unified with gravity \cite{swh76}. He has since renounced this
proposal and admits \textquotedblleft elementary quantum gravity
interactions do not lose information or quantum
coherence\textquotedblright\ \cite{swh05}. A second approach focuses
on the black hole remnant \cite{acn87}, stemmed from the idea of
correlation or entanglement of the radiation and the black hole. It
also fails because of the infinite degeneracy, which is hard to
reconcile with causality and unitarity \cite{jp92}. A third idea is
related to
 \textquotedblleft quantum hair\textquotedblright\ on a black hole \cite%
{kw89} that is found to be capable of store more information than one expects.
To resolve the paradox with this approach,
a projection onto local
quantum fields of low energies is required, and no one knows how this can be done.
A fourth approach is from Bekenstein \cite{jdb93} who suggests that
if the radiation spectrum is analyzed in detail,
enough non-thermal features might exist to encode all lost information.
Recently a new approach is
brought forward along the lines of quantum teleportation and the so-called
final state projection \cite{hm04}. The quantum information is estimated
to be capable of escaping with a fidelity of about $85\%$ on average \cite{sl06},
although whether the final state projection exists or not and, how it
can be justified,
remains a mystery. Finally, another recent work attracted serious attention after it
ruled out the possibility that information about the infallen matter could
hide in the correlations between the Hawking radiation and the internal
states of a black hole \cite{bp07}. The current state of affairs is a
direct confrontation: either unitarity or Hawking radiation being thermal
must break down.

In the original treatment, Hawking considered a fixed background geometry
without enforcing energy conservation \cite{swh74,swh75}. In contrast,
energy conservation is crucial in an improved treatment
by Parikh and Wilczek that considers s-wave outgoing particles,
or the Hawking radiation, as due to quantum tunneling,
and obtains a non-thermal spectrum for the Schwarzschild black hole \cite{pw00}.
The non-thermal probability distribution is
related directly to the change of entropy in a black hole \cite{pw00}.
In this Letter, we show that the non-thermal distribution
implies information can be coded into the correlations of sequential
emissions. We find that entropy remains conserved in the radiation
process; which leads naturally to the conclusion that
the process of Hawking radiation is unitary, and no
information loss occurs.
This implies that
even in a semiclassical treatment of the Hawking radiation process,
unitarity is not violated. The so-called black hole information
loss paradox arises from the neglect of energy conservation or
self-gravitational effect.

We start with a brief review of Hawking radiation as due to quantum
tunneling \cite{pw00}. Unlike the Schwarzchild coordinate,
the derivation that makes use of the Painlev\'{e}
coordinate system is regular at
the horizon and thus is particularly convenient for tunneling calculation.
Particles are supplied by considering the geometrical limit because of the
infinite blueshift of the outgoing wave-packets near the horizon. The
barrier is created by the outgoing particle itself, which is ensured by
energy conservation. The radial null geodesic motion is considered, and the
WKB approximation is adopted to arrive at the tunneling probability
\begin{eqnarray}
\Gamma &\sim & \exp \left[ -2\,\mathrm{Im}(I)\right]\nonumber\\
& =&\exp \left[ -8\pi E\left(
M-\frac{E}{2}\right) \right]\equiv\exp \left( \Delta S\right),  \label{tp}
\end{eqnarray}%
as the imaginary part of the action, and is
related to the change of the black hole's Bekenstein-Hawking entropy
on the second line, as was shown in Ref. \cite{pw00}.
This result Eq. (\ref{tp}) is clearly
distinct from the thermal distribution: $\Gamma (E)=\exp \left( -8\pi
EM\right) $, thus subsequent Hawking radiation emissions must be correlated
and capable of carrying away information encoded within. Further insight can
be gained if we compare with the general form of a quantum transition
probability \cite{mkp04}, expressed as
\begin{equation*}
\Gamma \sim \frac{e^{S_{\mathrm{final}}}}{e^{S_{\mathrm{initial}}}}=\exp
\left( \bigtriangleup S\right) ,
\end{equation*}%
in terms of the entropy change $\Delta S= S_f-S_i$
between the final and initial entropies $S_f$ and $S_i$.
This is in agreement with the tunneling probability, up to a factor
containing the square of the amplitude of the process. In other words,
the non-thermal Hawking radiation Eq. (\ref{tp}) reveals the
possibility of unitarity and no information loss.

We will find out whether or not there exist statistical correlations
between quanta of Hawking radiation. This was first discussed
in Refs. \cite{mkp04,amv05} by considering two emissions
with energies $E_{1}$ and $E_{2}$, or one emission
with energy $E_{1}+E_{2}$. The function
\begin{eqnarray}
C\left( E_{1}+E_{2;}\,E_{1},E_{2}\right) =\ln \Gamma \left(
E_{1}+E_{2}\right) -\ln [\Gamma \left( E_{1}\right) \Gamma \left(
E_{2}\right) ]\nonumber
 \label{cf}
\end{eqnarray}%
was used to measure statistical correlation between the two
emissions. With the following function forms
\begin{eqnarray}
\Gamma (E_{1}) &=& \exp \left[ -8\pi E_{1}\left( M-\frac{E_{1}}{2}\right) %
\right], \nonumber \\
\Gamma (E_{2}) &=& \exp \left[ -8\pi E_{2}\left( M-E_{1}-\frac{E_{2}}{2}%
\right) \right], \label{w1} \\
\Gamma (E_{1}+E_{2}) &=& \exp \left[ -8\pi (E_{1}+E_{2})\left( M-\frac{%
E_{1}+E_{2}}{2}\right) \right], \nonumber
\end{eqnarray}
$C\left( E_{1}+E_{2;}E_{1},E_{2}\right) =0$ is found, and
Refs. \cite{mkp04,amv05} wrongly conclude that no
correlation exists, including the
case of tunneling through a quantum horizon \cite{amv05}.
This makes no sense.
The notations used in the above (adopted from Refs. \cite{mkp04,amv05})
for $\Gamma (E_{1})$, $%
\Gamma (E_{2})$, and $\Gamma (E_{1}+E_2)$ are incorrect.
In particular, the form of the function $\Gamma (E_{2})$ [Eq. (\ref{w1})]
is misleading because it is different from Eq. (\ref{tp}).
To properly evaluate statistical correlation \cite{gs92}, it is
important to distinguish between statistical dependence or independence.
If the probability of two events arising simultaneously is identically the
same as the product probabilities of each event occurring independently,
these two events are independent or non-correlated. Otherwise, they are
dependent or correlated.
Because of the non-thermal nature, the probability $\Gamma \left(
E_{2}\right)$ used in Eq. (\ref{w1}) is not independent; instead, it is
conditioned on the emission with energy $E_{1}$.

The proper forms for the probabilities $\Gamma (E_{1})$ and $%
\Gamma (E_{2})$ are derived in the appendix using the
standard approach: $\Gamma (E_{1})=\int \Gamma (E_{1},E_{2})dE_{2}$ and $%
\Gamma (E_{2})=\int \Gamma (E_{1},E_{2})dE_{1}$, where the probability for
simultaneously two emissions with energies $E_1$ and $E_2$ is $\Gamma
(E_{1},E_{2})=\Gamma (E_{1}+E_{2})$. We find both independent
probabilities take the expected functional form of Eq. (\ref{tp}),
\begin{eqnarray}
\Gamma (E_{2}) =\exp \left[ -8\pi E_{2}\left( M-\frac{E_{2}}{2}\right) %
\right] ,  \label{2tp}
\end{eqnarray}
which then gives
\begin{eqnarray}
\ln \Gamma (E_{1}+E_{2})-\ln \left[ (\Gamma (E_{1})\ \Gamma (E_{2})\right]
=8\pi E_{1}E_{2}\neq 0,  \label{co}
\end{eqnarray}%
unlike what was concluded previously \cite{mkp04,amv05}.

Equation (\ref{co}) is the central result of this work.
To better understand its
implications we can make connection to a closely related topic in quantum
information.
Our result Eq. (\ref{co}) shows that subsequent emissions are
statistically dependent, and correlations must exist between them.
For sequential emissions of energies $%
E_{1}$ and $E_{2}$, the tunneling probability for the
second emission with energy $%
E_{2}$ should be understood as conditional probability given
the occurrence of tunneling of the particle with energy $E_{1}$. Thus,
instead of the misleading Eq. (\ref{w1}),
a proper notation is
\begin{eqnarray}
\Gamma (E_{2}|E_{1})=\exp \left[ -8\pi E_{2}\left( M-E_{1}-\frac{E_{2}}{2}%
\right) \right] ,  \label{cp}
\end{eqnarray}%
defined according to $\Gamma
(E_{1},E_{2})=\Gamma (E_{1})\cdot \Gamma (E_{2}|E_{1})$. The Bayesian law $%
\Gamma (E_{2}|E_{1}){\Gamma (E_{1})=\Gamma (E_{1}|E_{2}) \Gamma (E_{2})}$
then self-consistently connects between different probabilities.

Analogously, the conditional probability $\Gamma (E_{i}|E_{f})=\exp \left[
-8\pi E_{i}\left( M-E_{f}-\frac{E_{i}}{2}\right) \right] $ corresponds to
the tunneling probability of a particle with energy $E_{i}$ conditional on
radiations with a total energy of $E_f$. The entropy taken away by the
tunneling particle with energy $E_i$ after the black hole has emitted
particles with a total energy $E_{f}$ is then given by
\begin{eqnarray}
S\left( E_{i}|E_{f}\right) =-\ln \Gamma (E_{i}|E_{f}).  \label{te}
\end{eqnarray}%
In quantum information theory \cite{nc00}, $S\left(E_{i}|E_{f}\right)$
denotes conditional entropy, and it measures the entropy of $E_{i}$ given
that the values of all the emitted particles with a total energy $E_{f}$ are
known. Quantitatively, it is equal to the decrease of the entropy of a black
hole with mass $M-E_{f}$ upon the emission of a particle with energy $E_i$. Such
a result is consistent with the thermodynamic second law of a black hole \cite{swh760}:
the emitted particles must carry entropies
in order to balance the total entropy of the black hole and the
radiation. In what follows we show that the amount of correlation Eq. (\ref{co})
hidden inside Hawking radiation is precisely equal to mutual information.

The mutual information \cite{nc00} in a composite quantum system composed of
sub-systems $A$ and $B$ is defined as%
\begin{equation*}
S(A:B)\equiv S(A)+S(B)-S(A,B)=S(A)-S(A|B),
\end{equation*}%
where $S(A|B)$ is the conditional entropy. It is a legitimate measure for
the total amount of correlations between any bi-partite system. For
sequential emission of two particles with energies $E_{1}$ and $E_{2}$,
we find
\begin{eqnarray}
S(E_{2}:E_{1}) &\equiv & S(E_{2})-S(E_{2}|E_{1})\nonumber\\
&=& -\ln \Gamma (E_{2})+\ln \Gamma
(E_{2}|E_{1}).  \label{tmi}
\end{eqnarray}%
Using Eqs. (\ref{2tp}) and (\ref{cp}), we obtain $S(E_{2}:E_{1})=8\pi
E_{1}E_{2}$, \textit{i.e.}, the correlation of Eq. (\ref{co}) is
exactly equal to the mutual information between the two sequential
emissions.


We now count the entropy carried away by Hawking radiations. The
entropy of the first emission with an energy $E_{1}$ from a black
hole of mass $M$ is
\begin{eqnarray}
S(E_{1})=-\ln \Gamma (E_{1})=8\pi E_{1}\left( M-\frac{E_{1}}{2}\right).
\label{tp1}
\end{eqnarray}%
The conditional entropy of a second emission with an energy $E_{2}$ after
the $E_{1}$ emission is
\begin{eqnarray}
S(E_{2}|E_{1})=-\ln \Gamma (E_{2}|E_{1})=8\pi E_{2}\left( M-E_{1}-\frac{E_{2}%
}{2}\right) .  \label{ctp}
\end{eqnarray}%
The total entropy for the two emissions $E_{1}$ and $E_{2}$ then becomes
\begin{equation*}
S(E_{1},E_{2})=S(E_{1})+S(E_{2}|E_{1}),
\end{equation*}%
and the mass of the black hole reduces to $M-E_{1}-E_{2}$ while
it proceeds with the emission of energy $E_{3}$ with an entropy $%
S(E_{3}|E_{1},E_{2})=-\ln \Gamma (E_{3}|E_{1},E_{2})$. The total entropy of
three emissions at energies $E_{1} $, $E_{2}$, and $E_{3}$ is
\begin{equation*}
S(E_{1},E_{2},E_{3})=S(E_{1})+S(E_{2}|E_{1})+S(E_{3}|E_{1},E_{2}).
\end{equation*}%
Repeating the process until the black hole is completely exhausted, we find
\begin{eqnarray}
S(E_{1},E_{2},\cdots ,E_{n})=\sum\limits_{i=1}^{n}S(E_{i}|E_{1},E_{2},\cdots
,E_{i-1}),  \label{bhe}
\end{eqnarray}%
where $M=\sum_{i=1}^{n}E_{i}$ equals to the initial black hole mass
due to energy conservation and $S(E_{1},E_{2},...,E_{n})$ denotes
the joint entropy of all emissions while $S(E_{i}|E_{1},E_{2},\cdots
,E_{i-1}) $ is the conditional entropy. Equation (\ref{bhe}) then
corresponds to nothing but the chain rule of conditional entropies
in quantum information theory \cite{nc00}. In the appendix, we find
the total entropy $S(E_{1},E_{2},...,E_{n})=4\pi M^{2}$ exactly
equals the black hole's Bekenstein-Hawking entropy. This result is
independently verified by counting of microstates of Hawking
radiations as shown in the appendix.

The reason information can be carried away by black hole radiation is
the probabilistic nature of the emission itself. Given the emission rate $%
\Gamma (E)\sim \exp \left[ -8\pi E\left( M-\frac{E}{2}\right) \right] $, one
knows definitively that a radiation of energy $E$ may occur with a
probability $\Gamma (E)$. In other words, the uncertainty of the event (an
emission with an energy $E$) or the information we can gain, on
average, from the event is $S(E)=-\ln \Gamma (E)$. When an emission with an
energy $E_{1}$ is received, the potential gain in information is $%
S(E_{1})=-\ln \Gamma (E_{1})$. When the next emission with an
energy $E_{2}$ is received, an additional information $S(E_{2}|E_{1})=-\ln
\Gamma (E_{2}|E_{1})$ can be gained, which is conditional on already
receiving the emission of an energy $E_{1}$. Continuing on, we compute the
information gained from all emissions until the black hole
is exhausted. The total entropy carried out by radiations is then found to
be $S(E_{1},E_{2},...,E_{n})=4\pi M^{2}$, which means all the entropy of the
black hole is taken out by its Hawking radiations. Putting together our
earlier result that the entropy carried away by an emission is
the same as the entropy reduction of the accompanying black hole
during each emission, we conclusively show that entropies of Hawking
radiations and their accompanying black holes are conserved during black
hole radiation. According to quantum mechanics, a unitary process does not
change the entropy of a closed system. This implies that the process
of Hawking radiation
is unitary in principle, and no information loss is expected.


In conclusion, through a careful reexamination of Hawking radiation, we
discover and quantify correlations amongst radiated particles in terms of
Eq. (\ref{co}). Our result for the first time provides a clear picture of
how and how much information can be carried away by Hawking radiation from a
black hole. Although the prospect for information hidden inside Hawking
radiation has been discussed time and again, earlier works do not enforce
energy conservation strictly and assumed a thermal distribution for the
radiated particles (please see \cite{jp92} and references therein). In
contrast, our study is built on the principle of energy conservation, where
the effect of self-gravitation plays a crucial role, and the spectrum of
radiated particles is non-thermal. Making connection with information
theory, we find that entropy is strictly conserved during Hawking radiation,
\textit{i.e.}, the entropy of a black hole is coded completely in the
correlations of the emitted radiations upon its exhaustion.

Our conclusions show the information is
not lost, and unitarity is held in the process of Hawking radiation
although based on results within a semiclassical treatment for s-wave emissions
where energy conservation is enforced \cite{pw00}. For more elaborate
treatments, {\it e.g.}, those involving coding information in the correlations,
a complete quantum gravity theory may still be needed. However, our
analysis confirms that the energy conservation or self-gravitational
effect remains crucial for approaches
based on self-consistent quantum gravity theories.

Finally, we hope to point out that our analysis can be extended to
charged black holes, Kerr black holes, and Kerr-Neumann black holes.
Even for the situations involving quantum gravity effects or the
noncommutative black holes, our method remains effective in
providing consistent resolutions \cite{zcyz}. We show that due to
self-gravitation effect, information can come out in the form of
correlated emissions from a black hole, and our work thus resolves
the black hole information loss paradox.


This work is supported by National Basic Research Program of China
(NBRPC) under Grant No. 2006CB921203.

\section{appendix}

This appendix contains some details for
a few key steps supporting our
results as given in the main text.

The joint probability distribution of two simultaneous emissions
of energies $E_{1}$ and $E_{2}$ is
\begin{eqnarray}
\Gamma (E_{1},E_{2}) &=& \Gamma (E_{1}+E_{2})\nonumber\\
&=&\exp \left[ -8\pi
(E_{1}+E_{2})\left( M-\frac{E_{1}+E_{2}}{2}\right) \right],\nonumber
\end{eqnarray}
subjected to a normalization factor $\Lambda$, determined by
$\Lambda\int_{0}^{M}\exp [-8\pi E(M-\frac{E}{2})]dE=1$. The
independent probability distributions for a single emission $\Gamma (E_{1})$
or $\Gamma (E_{2})$ are $\Gamma (E_{1})=\Lambda
\int_{0}^{M-E_{1}}\Gamma (E_{1},E_{2})dE_{2}=\exp [-8\pi E_{1}(M-\frac{E_{1}%
}{2})]$ and $\Gamma (E_{2})=\Lambda \int_{0}^{M-E_{2}}\Gamma
(E_{1},E_{2})dE_{1}=\exp [-8\pi E_{2}(M-\frac{E_{2}}{2})]$ and are
identical in their function forms.
In the main text, our result Eq. (\ref{te}) reveals that Hawking
radiations are correlated and carry away that much entropy
from the black hole. We now show that the initial entropy
of a black hole is the same as the entropy of all emitted radiations upon
its exhaustion.

Assuming the tunneling/emission probability is given by Eq. (\ref{tp}),
when the black hole is
exhausted due to emissions, we can find the entropy of our system by counting the number of
its microstates. For example, one of the microstates is $\left(
E_{1},E_{2},\cdots ,E_{n}\right) $ and $\sum_{i}E_{i}=M$. Within such a
description, the order of $E_{i}$ cannot be changed, the distribution of
each $E_{i}$ is consistent with the discussion in the
main text. The probability for the specific microstate $\left(
E_{1},E_{2},\cdots ,E_{n}\right) $ to occur is given by
\begin{equation*}
P=\Gamma (M;E_{1})\times \Gamma (M-E_{1};E_{2})\times \cdots \times \Gamma
(M-\sum_{j=1}^{n-1}E_{j};E_{n}),
\end{equation*}%
with
\begin{eqnarray}
\Gamma (M;E_{1}) &=&\exp \left[ -8\pi E_{1}(M-E_{1}/2)\right] ,  \notag \\
\Gamma (M-E_{1};E_{2}) &=&\exp \left[ -8\pi E_{2}(M-E_{1}-E_{2}/2)\right] ,
\notag \\
&&\cdots ,  \notag \\
\Gamma (M-\sum_{j=1}^{n-1}E_{j};E_{n}) &=&\exp \left[ -8\pi
E_{n}(M-\sum_{j=1}^{n-1}E_{j}-E_{n}/2)\right] \nonumber\\
&=&\exp (-4\pi E_{n}^{2}),
\notag
\end{eqnarray}%
where $\Gamma (M;E_{1})$ denotes the probability Eq. (\ref{tp})
for a emission with energy $E_{1}$ by a black hole with mass $M$. Proceeding
with a detailed calculation, we find that $P=\exp (-4\pi M^{2})=\exp (-S_{%
\mathrm{BH}})$, where $S_{\mathrm{BH}}$ is the entropy of the black hole.
According to the fundamental postulate of statistical mechanics that all
microstates of an isolated system are equally likely, we find the number of
microstates $\Omega =\frac{1}{p}=\exp (S_{\mathrm{BH}})$. On the other hand,
according to the Boltzmann's definition, the entropy of a system is given by
$S=\ln \Omega =S_{\mathrm{BH}}$, (where the Boltzmann constant $k=1$ is taken.)
 Thus we prove that after a black hole is exhausted due to
Hawking radiation, the entropy carried away by all emissions
is precisely equal to the entropy in the original black hole.

\end{document}